\documentclass{ws-procs975x65}
\usepackage{mcite}
\usepackage{amssymb} 
\usepackage{amsmath} 
\usepackage{amsxtra} 
\usepackage{verbatim}

\newcommand{\GeV}{${\rm GeV}$}

\begin{document}

\title{
THE QCD COUPLING AND PARTON DISTRIBUTIONS\\ AT HIGH 
PRECISION
\footnote{Dedicated to M. Gell--Mann on the occasion of his 80th birthday.}
}

\author{JOHANNES BL\"UMLEIN%
}

\address{Deutsches Elektronen-Synchrotron, DESY,
Platanenallee 6, D-15738 Zeuthen, Germany\\
E-mail: Johannes.Bluemlein@desy.de}

\begin{abstract}
A survey is given on the present status of the nucleon parton distributions and related precision
calculations and precision measurements of the strong coupling constant $\alpha_s(M_Z^2)$. We also 
discuss the impact of these quantities on precision observables at hadron colliders.
\end{abstract}


\bodymatter

\section{Inside Nucleons}\label{sec1}

\vspace*{1mm}\noindent
The physics of the strong interactions always has been tightly connected to the study
of nucleons at shorter and shorter distances. The measurement of the anomalous magnetic moments
of the proton~\cite{Frisch:1933} and neutron~\cite{Alvarez:1940zz} in 1933 and 1939 
made clear that nucleons are no elementary particles. During the 1950ies the Hofstadter 
experiments~\cite{Hofstadter:1963} revealed the charge distributions inside 
nucleons~\cite{Schopper:1961} at scales $Q^2 \simeq 0.5 \cdot M_N^2$. Yet it was unknown how these 
distributions came about. In 1964 Murray Gell--Mann~\cite{GellMann:1964nj} proposed the quark model,
to catalog the plethora of observed baryons and mesons. Independently G. Zweig suggested
aces~\cite{Zweig:1964jf} as the building blocks of hadrons. A direct connection to the lepton-nucleon 
scattering data was not made at that time. 

Back in 1954 C.N. Yang and R Mills~\cite{Yang:1954ek}
proposed novel bosonic field theories based on gauge invariance with respect to non-abelian groups.
This development went unrelated to strong interactions for a long time. With the advent of the 
Stanford Linear Accelerator in 1968 the nucleon structure could be resolved at much shorter distances
by the MIT-SLAC 
experiments~\cite{Panofsky:1968pb,Taylor:1969xi,Kendall:1991np,*Taylor:1991ew,*Friedman:1991nq} beyond 
the resonant region $W \geq 2 {\rm GeV}$ for values $Q^2$ up to $30~{\rm GeV}^2$. The remarkable 
finding by 
these experiments were that $i)$ the structure function $\nu W_2(\nu,Q^2)$ which has been expected
to depend on both kinematic variables $\nu$ and $Q^2$ independently, turned out to take the 
same values for fixed values of $ x = Q^2/(2 M_N \nu)$ irrespectively of $\nu$ and $Q^2$ at high enough 
values. This phenomenon is called {\sf scaling}. 
$ii)$ The ratio of the longitudinal structure function $W_L$ and $W_2$ turned out to be 
very small. Bjorken~\cite{Bjorken:1969} had predicted scaling at asymptotic scales $Q^2, \nu 
\rightarrow \infty$ in 1969. Learning about the SLAC-MIT results R.~Feynman very quickly proposed
the parton {model~\cite{Feynman:1969wa,*Feynman:1973xc}}, which is equivalent to Bjorken's description 
but based 
on the {\sf observed} strict microscopic correlation between $Q^2$ and $\nu = q.p_i$
\begin{eqnarray}
W(x,Q^2) = \sum_i e_i^2 \int_0^1 d x_i f_i(x_i) \delta\left(\frac{q.p_i}{M^2} - 
\frac{Q^2}{M^2}\right)~, 
\end{eqnarray}
where $e_i$ and $f_i$ denote the parton's charge and distribution functions. Would the parton model
be unique in describing the new data? This has been challenged by other popular formalisms like vector 
meson {dominance~\cite{Sakurai:1969}}. However, they failed to describe the behaviour observed for 
$W_L$, which corresponded to that of spin 1/2 partons, according to the calculations by Callan and 
{Gross~\cite{Callan:1969uq}}. 

Yang--Mills theories~\cite{Yang:1954ek} became building blocks of the electro-weak Standard 
{Model~\cite{Glashow:1961tr,*Weinberg:1967tq}}, although there renormalizibility had not been
proven yet, a conditio sine qua non for a physical theory. The proof was an urgent matter and in 
1971 it was achieved both for massless and spontaneously broken Yang-Mills theories, along
with designing practical loop computations in this sophisticated theory in an automated 
{way~\cite{'tHooft:1971fh,*'tHooft:1972fi,*'tHooft:1972ue,*'tHooft:1973mm,*'tHooft:1973pz,*'tHooft:1978xw,
          Veltman:1963}}.
Quantum Chromodynamics (QCD) was proposed as the theory of the strong interactions in 1972 by
Gell--Mann and Fritzsch and Leutwyler~\cite{Fritzsch:1972jv,Fritzsch:1973pi} as a renormalized 
Yang-Mills field theory based on $SU(3)$ gauge {interactions~\cite{Nambu:1966}}.
D.~Gross, F. Wilczek~\cite{Gross:1973id}   and D. Politzer~\cite{Politzer:1973fx}
studied the asymptotic behaviour of color octet 
gluon Yang-Mills theory, cf. {also~\cite{Khriplovich:1969aa,*tHooft:unpub}}, and found asymptotic 
freedom. This is the essential ingredient, which makes it possible to perform perturbative
calculations at large scales in a theory with strong interactions at low scales.  

At short distances the nucleon structure functions $F_i(x,Q^2)$ obey the light-cone 
{expansion~\cite{Wilson:1969zs,
*Brandt:1970kg,*Frishman:1971qn}}. At large scales $Q^2$ the contributions of lowest twist dominate 
and the representation
\begin{eqnarray}
F_i(x,Q^2) = \sum_j C_{i}^j(x,Q^2/\mu^2) \otimes f_j(x,\mu^2)
\label{eq:1}
\end{eqnarray}
holds. Here $C_{i}^j(x,Q^2/\mu^2)$ denote the Wilson coefficients and $f_j(x,\mu^2)$ are the
parton densities. $\mu^2$ is an arbitrary factorization scale and $\otimes$ denotes the Mellin 
convolution.

The scale behavior of the nucleon structure functions $F_i(x,Q^2)$ obey renormalization group 
equations, an important aspect of renormalizable Quantum Field Theories to which Murray Gell--Mann made
very essential contributions very {early~\cite{GellMann:1954fq}\footnote{It is interesting to note 
that different approaches to renormalization result into different mathematical 
structures as shown {in~\cite{Brunetti:2009qc}}.Thus the 
method by Gell--Mann and 
Low~\cite{GellMann:1954fq} is in general related to a cocycle, while that by  
St\"uckelberg and Petermann~\cite{Stuckelberg:1951gg} relates to a group.
I thank A. Petermann for pointing out Ref.~\cite{Brunetti:2009qc} to me.}}.
Transforming Eq.~(\ref{eq:1}) to Mellin space one obtains the following 
Callan-Symanzik~\cite{Callan:1970yg,*Symanzik:1970rt} equations~:
\begin{eqnarray}
\left[
\mu \frac{\partial}{\partial \mu} + \beta(g) \frac{\partial}{\partial g} - 2 \gamma_\psi(g) \right]
F_i(N,Q^2) &=& 0 
\end{eqnarray}
\begin{eqnarray}
\left[
\mu \frac{\partial}{\partial \mu} + \beta(g) \frac{\partial}{\partial g} + \gamma_\kappa(N,\mu)
- 2 \gamma_\psi(g) \right]
f_k(N,\mu^2) &=& 0 \\
\left[
\mu \frac{\partial}{\partial \mu} + \beta(g) \frac{\partial}{\partial g} - \gamma_\kappa(N,\mu)\right]
C_j^k(N,Q^2/\mu^2) &=& 0~, 
\end{eqnarray}
with $\beta(g)$ the QCD $\beta$-function and $ \gamma_\kappa(N,\mu)$ the anomalous dimensions. Both 
functions imply the scaling violations of the structure functions.\footnote{We remind that Drell and
collaborators at the end of the 1960ies~\cite{Drell:1969jm,*Drell:1969wb,*Drell:1969wd,
*Yan:1969gn,*Drell:1970yt} were seeking desperately scaling in fermion-meson 
interactions with loop corrections but ended up with scaling violations in general.}
With progressing time the measurement of the deep-inelastic structure functions improved considerably
and after 40 years, e.g. the precision  of the structure function $F_2^{\rm em}(x,Q^2)$ reached 1\%
over a very wide range, {cf. \cite{H1Z:2009wt}}. Due to this both the precision measurement of the 
unpolarized 
parton distributions and the strong coupling constant $\alpha_s(M_Z^2)$ is possible from these data.
\section{Higher Order QCD Corrections to Deep-Inelastic 
         Scattering}\label{sec2}

\vspace*{1mm}\noindent
On the theory side, the progress in higher order computations, likewise, has been enormous
during the same period. The initial 1-loop 
results~\cite{Gross:1973id,Politzer:1973fx,
Gross:1973ju,*Gross:1974cs,*Georgi:1974sr,*Furmanski:1981cw}
are now widely improved to 3-loop order 
and even somewhat beyond. This is necessary to  comply with the current precision of data. 
The status of the theory of deep-inelastic scattering is illustrated in the flowchart below.
Here the dates indicate the year in which the corresponding correction to the 
respective quantities has been calculated. 

Let me mention the most far reaching results. For the QCD $\beta$-function the 4--loop corrections
were first computed by Vermaseren et al.~\cite{Vermaseren:1997fq} in 1997. The unpolarized anomalous 
dimensions and massless Wilson coefficients are known to 
3--loops for a series of 
moments~\cite{Larin:1993vu,*Larin:1996wd,*Retey:2000nq,*Blumlein:2004xt} and in complete 
form~\cite{Moch:2002sn,Vogt:2004mw,Vermaseren:2005qc}
since 2004/05.
A first moment for the non-singlet+ anomalous dimensions has been computed at 
4--loops~\cite{Baikov:2006ai}
in 2006 and more moments are in preparation. The heavy flavor Wilson coefficients in the region
$Q^2/m^2 = \rho \gg 1$, which is a good approximation in case of $F_2(x,Q^2)$ for $\rho \geq 10$,
were computed for a larger number of moments~\cite{Bierenbaum:2009mv} and in complete form {for 
$F_L$~\cite{Blumlein:2006mh}}.
Currently the computation of the Wilson coefficients at general values of $N$ is 
{underway~\cite{Ablinger:2010ha}}.

At the level of the leading twist $(\tau = 2)$ representation the light cone expansion and the 
QCD-improved parton model lead to the same results. In the 1980ies it was thought, that
the higher order corrections need to be supplemented by different small-$x$ 
resummations, {cf.~\cite{Fadin:1975cb,Gribov:1981ac}},
to obtain correct results, even in the HERA kinematic region. These perturbative resummations are 
connected to the problem how, within their approach, perturbative and non-perturbative contributions
are clearly separated - an important pre-requisite to apply perturbation theory at all. 

\vspace*{6mm}
\begin{center}
\begin{figure}[h]
\hspace*{1cm}
\psfig{file=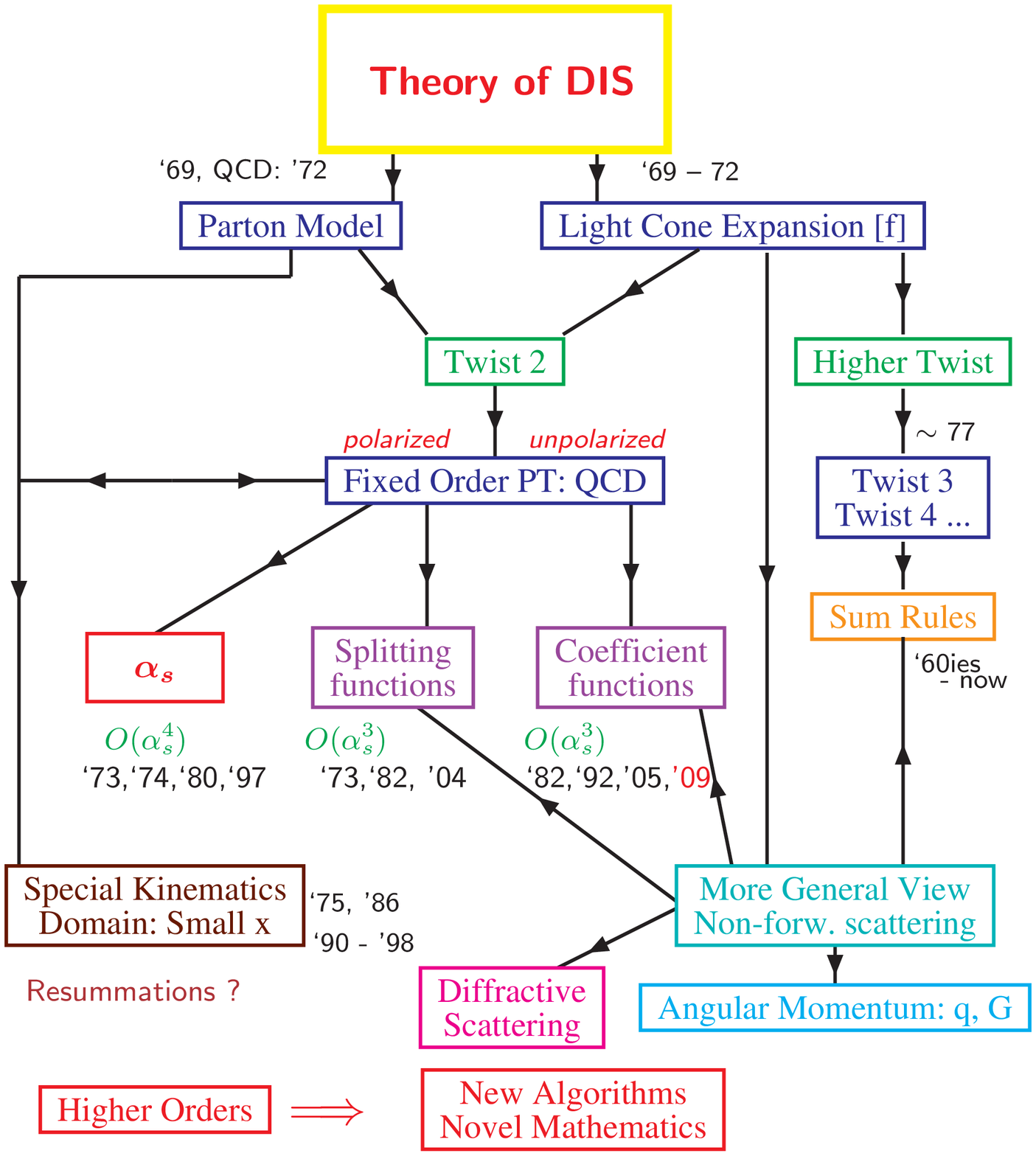,width=3.5in}
\end{figure}
\end{center}
The resummation
\cite{Fadin:1975cb,Fadin:1998py} has successfully predicted the so-called `leading' poles of the QCD 
anomalous 
dimensions related to the poles at $N = 1$ in Mellin space, at least up to $O(\alpha_s^3)$. The leading 
series is related to the scale-invariant limit of QCD. The corresponding resummed anomalous dimension, 
however, has branch cuts in the complex plane \cite{Ellis:1995gv,*Blumlein:1995mi} but no poles at all, see 
{also~\cite{Blumlein:1998mg}}. 
These singularities are
much milder. Phenomenological 
studies \cite{Blumlein:1995jpxBlumlein:1996ddxBlumlein:1996hbxBlumlein:1997em,*Blumlein:1998pp} have 
shown, that 
subleading effects are as important as
the leading ones, since they widely cancel the effect of the former. One estimates that about four 
complete series of these terms are needed to obtain convergence. Currently the only practical approach
relies on the computation of the Wilson-coefficients and anomalous dimensions to high enough order,
which includes all the small- and large-$x$ effects automatically. In the latter case, the 
renormalization group even allows reliable {resummations~\cite{Sterman:1995fz,*Laenen:2008gt}}.

Beyond the level of leading twist much less is known on deeply-inelastic structure functions. Most of the 
results obtained so far concern the 1-loop level, cf. {e.g.~\cite{Braun:2009vc}}\footnote{For older 
references {see~\cite{Blumlein:1998nv}}.}. Here, the corresponding 
partonic operator matrix elements depend on several dimensionless invariants $x_i$, unlike in the case
of lowest twist. They cannot be measured individually in the deep-inelastic process, but require 
ab-initio determinations using reliable non-perturbative methods. For a series of moments, this may 
be possible in the future, using lattice techniques. In the polarized case a series of results has been 
obtained for the twist-3 contributions, 
{cf. e.g.~\cite{Kodaira:1994ge,*Geyer:1996isa,*Kodaira:1999ih,*Ji:2000ny,*Belitsky:2000pb,*Braun:2000yi,
*Braun:2001qx,*Braun:2009mi}}, 
among them a series of integral relations between 
different structure {functions~\cite{Blumlein:1996vs,Blumlein:1998nv}}. Finally, deep-inelastic 
non-forward scattering has been
extensively studied during the last two {decades, \cite{Belitsky:2005qn}}. These methods do in principle 
allow to 
measure the (quark) angular momentum of {nucleons \cite{Ji:1996ek}}, which is still difficult 
experimentally. One may apply
these techniques to describe inclusive deep-inelastic diffractive scattering, referring to the 
light cone expansion \cite{Blumlein:1999sc,*Blumlein:2001xf,*Blumlein:2002fw} and proving that the anomalous 
dimensions are structurally the same as in the forward case.  

\vspace*{-4mm}
\section{Precision Quark and Gluon Twist-2 Distributions}\label{sec3}

\vspace*{1mm}\noindent
In the following we describe recent extractions of the unpolarized twist-2
parton densities to 3-loop accuracy and higher.
\begin{center} \begin{figure}[t]\vspace*{6mm} \hspace*{2cm} 
\psfig{file=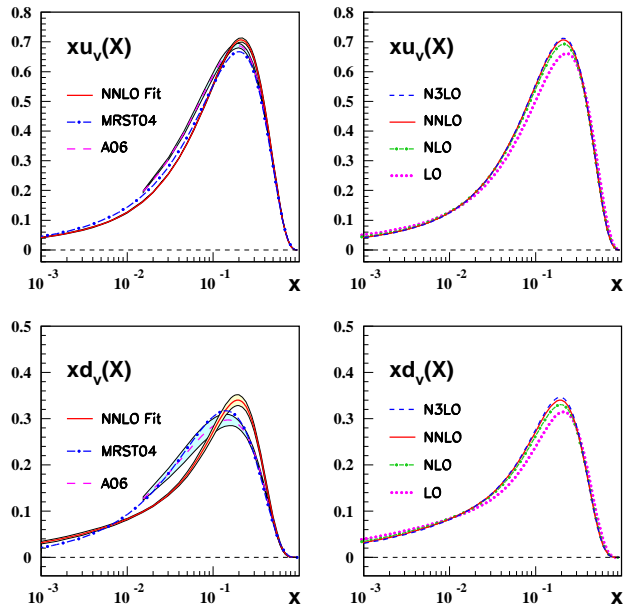,width=3.2in} 
\caption[]{Left panels: The parton densities $xu_v$ and $xd_v$ at 
the input scale $Q_0^2 = 4.0~\GeV^2$ (solid line) compared to results obtained from NNLO analyses by MRST04 
(dashed--dotted line)~\cite{Martin:2004ir} and A06 (dashed line)~\cite{Alekhin:2006zm}. The shaded areas 
represent the fully correlated $1\sigma$ statistical error bands. Right panels: Comparison of the same 
parton densities at different orders in QCD from LO to N$^3$LO {Ref.~\cite{ Blumlein:2006be}}.} 
\label{aba:fig1} \end{figure} 
\end{center} 

\vspace*{-6mm} 
\subsection{Flavor non-singlet analysis}\label{sec31}
The flavor non-singlet parton distributions obey scalar evolution equations
and do 
not depend on the gluon distribution, which is more difficult to
access in deep-inelastic scattering and may cause some systematic uncertainty, in particular determining the 
strong coupling constant $\alpha_s(M_Z^2)$. Moreover, in the small $x$ region the QCD evolution leads to 
moderate changes of the the distributions, unlike in the flavor singlet and gluon-case. Due to this flavor 
non-singlet analyses are advantageous. One may apply the valence-approximation in the region $x \geq 
x_0,~~x_0 \sim 0.35, 0.4$ and construct a non-singlet distribution from deuteron and proton data for $x \leq 
x_0$. To describe the valence quark parton densities, the distribution $x(\overline{d} - 
\overline{u})(x,Q^2)$ has to be known, which can be measured using Drell-Yan data. Furthermore, the 
non-singlet $O(\alpha_s^2)$ heavy flavor corrections are applied, which amount to about $1\%$. The yet 
unknown 3-loop corrections will be even smaller. To perform a leading twist analysis, kinematic regions with 
higher twist contributions are cut out in a systematic study, implying the cuts of $Q^2 > 4~\GeV^2, W^2 > 
12.5~\GeV^2$, {cf.~\cite{Blumlein:2004ip,Blumlein:2006be}}. The results for the parton distributions 
$xu_v(x,Q^2)$ and $xd_v(x,Q^2)$ are illustrated in Figure~1 and are compared to other determinations. Wile 
in 
the case of the $xu_v$ distributions the overall agreement is good, there are still systematic differences 
in case of the down-valence distribution. We also illustrate the perturbative expansion from LO to N$^3$LO 
reaching convergence.

Extrapolating the twist-2 QCD fit results into the region $12.5~\GeV^2  > 
W^2 > 4~\GeV^2$ the flavor non-singlet higher twist contributions can be 
determined {empirically~\cite{Blumlein:2008kz}}. The inclusion of soft resummation
terms for the Wilson coefficient beyond the N$^3$LO corrections allows
to extract the higher twist terms in the region $x < 0.75$ in a stable way,
while lower order analyses overestimate the higher twist contributions. The
relative higher twist contributions in the proton and deuteron case turn out
to be about of the same size. 

\vspace*{-5mm}
\subsection{Combined singlet and non-singlet analysis}\label{sec32}

\vspace*{1mm}\noindent
In combined singlet and non-singlet analyses of the deep-inelastic world data at NNLO, {cf. 
Refs.~\cite{Alekhin:2009vn,Alekhin:2010iu,JimenezDelgado:2008hf,JimenezDelgado:2009tv,Martin:2009iq}}, one 
determines also the different sea-quark and gluon densities. This has always to be done together with the
measurement of the QCD scale $\Lambda_{\rm QCD}$ due to strong correlations. To unfold the sea-quark 
densities one refers to Drell-Yan- and di-muon data as well, through which the distributions
$x(\overline{d}-\overline{u})(x,Q^2)$ and $xs(x,Q^2)=x\overline{s}(x,Q^2)$ can be measured individually.
Due to the large charm-quark contribution to the deep-inelastic structure functions the description
of the heavy flavor contributions is required at the same level of accuracy as for the light 
partons. Currently it is available to $O(\alpha_s^2)$~\cite{Laenen:1992zkxLaenen:1992xs,*Riemersma:1994hv} 
and the $O(\alpha_s^3)$ corrections are 
{underway~\cite{Bierenbaum:2008yu,Bierenbaum:2009mv,Ablinger:2010ha}}.
To obtain an accurate interpolation between low and higher scales $Q^2$, the so-called BMSN-interpolation
is {recommended~\cite{Buza:1996wv,Bierenbaum:2009zt}}. Nowhere in the kinematic region of HERA heavy flavor
logs become large to be {resummed~\cite{Gluck:1993dpa}}, i.e. in a very wide kinematic region even the charm 
quarks cannot be viewed massless. Yet, one may {\sf define} heavy flavor parton densities in terms of
technical quantities~\cite{Buza:1996wv,Bierenbaum:2009mv} to some extent~\footnote{At 3-loop order graphs 
exist with both charm and bottom-quark lines in the operator matrix elements. They do not fall under 
the paradigm of single parton distributions,despite being universal.} to evaluate other observables. Here 
one has always to check
to which extent this approximation holds. 

The present analyses have lead to very precise parton distributions. 
The NNLO parton distributions determined in Refs.~\cite{JimenezDelgado:2008hf,Alekhin:2009vn} do widely
agree within the measured region, with very slight differences in the $x(u+\overline{u})$ and  
$x(d+\overline{d})$ distributions.  In Figure~2 we compare the results of the NNLO fits of 
{Refs.~\cite{Martin:2009iq,Alekhin:2009vn}} for the light partons. At low scales $\mu^2$ the 
sea-quark and gluon distributions~\cite{Martin:2009iq} take lower values than those in 
{Ref.~\cite{Alekhin:2009vn}}, and the NNLO gluon distribution even tends to negative values yielding
the largest difference. 

Through the evolution the densities get closer. This, however, is partly due to the
large value of $\alpha_s(M_Z^2)$, Eq.~(\ref{eq:aM}), which leads to a relative acceleration of the evolution
compared {to~\cite{Alekhin:2009vn}}. At scales larger than $Q^2 \sim 10^4~\GeV^2$, accessible at the LHC,
this will lead to a further growing gluon density of~~\cite{Martin:2009iq} compared 
{to~\cite{Alekhin:2009vn}}.
The precision observables at the LHC will help to constrain the parton distributions further.
\begin{center}
\begin{figure}[t]
\hspace*{.3cm}
\psfig{file=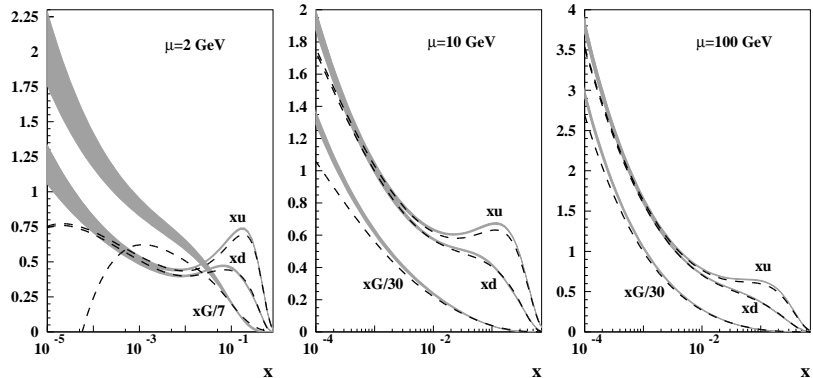,width=4.5in}
\caption[]{The light parton densities $xu, xd$ and $xG$ at the scales  $\mu^2 = 4, 100, 10000~\GeV^2$.
The bands denote the parton distributions with  1$\sigma$ uncertainty of {ABKM09~\cite{Alekhin:2009vn}}. 
The dashed lines correspond to {MSTW08~\cite{Martin:2009iq}}; from {Ref.~\cite{Alekhin:2009vn}}.
}
\label{aba:fig11}
\end{figure}
\end{center}
\noindent

\vspace*{-5mm}
\section{The Strong Coupling Constant}\label{sec4}

\vspace*{1mm}\noindent
The QCD parameter $\Lambda_{\rm QCD}$, or $\alpha_s(M_Z^2)$, is determined 
in QCD fits together with the non-perturbative input densities for the
different partons at a starting scale $Q^2_0$. There are tight correlations
between the value of $\alpha_s(M_Z^2)$ and some parameters of the parton densities.
An important example is the normalization of the gluon density, {cf.~\cite{Alekhin:2009vn,Blumlein:2010rn}}.
In the non-singlet analysis \cite{Blumlein:2006be} we obtained at NNLO, cf. {also~\cite{Gluck:2006yz}},
\begin{eqnarray}
\Lambda_{\rm QCD}^{N_f = 4} = 226 \pm 25~{\rm MeV}~,
\end{eqnarray}
and at N$^3$LO, assigning to the yet unknown 4-loop anomalous dimension a $\pm 100\%$ error,
\vspace*{-3mm}
\begin{eqnarray}
\Lambda_{\rm QCD}^{N_f = 4} = 234 \pm 26~{\rm MeV}~.
\end{eqnarray}
Usually the QCD parameter is expressed in terms of $\alpha_s(M_Z^2)$. In the following we
compare the results of recent NNLO and N$^3$LO analyses for the deep-inelastic world data
obtained by different groups~: 
\begin{alignat}{2}
\label{eq:al1}
\alpha_s(M_Z^2)  &= 0.1134 
                           {
\begin{array}{l} + 0.0019 \\
                                            - 0.0021 \end{array}} 
\hspace*{5.5mm} 
{\rm NNLO}  
&&\hspace*{5mm} \text{[\citen{Blumlein:2006be}]} 
\\
\label{eq:al1a}
\alpha_s(M_Z^2)  &= 0.1141 {
                           \begin{array}{l} + 0.0020 \\
                                            - 0.0022 \end{array}} 
\hspace*{5.5mm} 
{\rm N^3LO}   && 
\hspace*{5mm}
\text{[\citen{Blumlein:2006be}]} 
\\
\label{eq:al1b}
\alpha_s(M_Z^2)  &= 0.1135 \pm 0.0014 \hspace*{5mm} 
{\rm NNLO,~FFS}                           && \hspace*{5mm}
\text{[\citen{Alekhin:2009vn}]} 
\\
\label{eq:al2}
\alpha_s(M_Z^2)  &= 0.1129 \pm 0.0014 \hspace*{5mm} 
{\rm NNLO,~BSMN}                          && 
\hspace*{5mm}
\text{[\citen{Alekhin:2009vn}]} 
\\
\alpha_s(M_Z^2)  &= 0.1124 \pm 0.0020 \hspace*{5mm} 
{\rm NNLO,~dyn.~approach}                 && 
\hspace*{5mm}
\text{[\citen{JimenezDelgado:2008hf}]} 
\\
\alpha_s(M_Z^2)  &= 0.1158 \pm 0.0035 \hspace*{5mm} 
{\rm NNLO,~stand.~approach}               && 
\hspace*{5mm}
\text{[\citen{JimenezDelgado:2008hf}]} 
\\
\label{eq:aM}
\alpha_s(M_Z^2)  &= 0.1171 \pm 0.0014 \hspace*{5mm} 
{\rm NNLO}                                && 
\hspace*{5mm}
\text{[\citen{Martin:2009bu}]} 
\end{alignat}
More recent unpolarized NNLO fits, including the combined HERA {data~\cite{H1Z:2009wt}}, yield
\begin{alignat}{2}
\label{eq:al3}
\alpha_s(M_Z^2)  &= 0.1147 \pm 0.0012 
~~~{\rm NNLO}               && 
\hspace*{5mm}
\text{[\citen{Alekhin:2010iu}]} 
\\
\alpha_s(M_Z^2)  &= 0.1145 \pm 0.0042 
~~~{\rm NNLO,~preliminary}  && 
\hspace*{5mm}
\text{[\citen{H1ZPDF}]} 
\end{alignat}
Note that the values (\ref{eq:al1},\ref{eq:al1a}) are independent of the gluon 
distribution. The combined singlet non-singlet analysis, based on rather different data,
and being sensitive to both the sea-quark and gluon densities, yield the very similar values
(\ref{eq:al1b},\ref{eq:al2}). This analysis has been performed with a different code than used 
{in~\cite{Blumlein:2006be}}.

The above values are located below the present weighted average of $\alpha_s(M_Z^2)$ measurements 
\cite{Bethke:2009jm} of
\begin{eqnarray}
\label{alpWA}
\alpha_s(M_Z^2)  = 0.1184  \pm 0.0007~,
\end{eqnarray}
cf.~Figure~3. The error given in (\ref{alpWA}) cannot include the  yet unknown relative systematics 
between the different classes of the same type of measurement. 

We would like to mention that recent determinations of $\alpha_s(M_Z^2)$ using event shape
moments for high energy $e^+e^-$ annihilation data from PETRA and LEP including power corrections
the following values were obtained~:
\begin{alignat}{2}
\alpha_s(M_Z^2)  &= 0.1135 \pm 0.0002~{\rm (exp)} \pm 0.005~{(\Omega_1)} \pm 0.0009~{\rm (pert)} 
&&~{\rm NNLO}~~~
\text{[\citen{Abbate:2010vw}]}
\\
\alpha_s(M_Z^2)  &= 0.1153 \pm 0.0017~{\rm (exp)} \pm 0.0023~{\rm (th)} &&~{\rm NNLO}~~~    
\text{[\citen{Gehrmann:2009eh}]}.
\end{alignat}
Also these measurements of $\alpha_s(M_Z^2)$ yield low values. They show that the results
obtained analyzing deep--inelastic data do not form a special case. Also in deep-inelastic scattering
off polarized targets $\alpha_s(M_Z^2)$, at NLO, has been measured, however with larger errors,
{see~\cite{Blumlein:2010rn,
Blumlein:2010ri,Bluemlein:2002be}}. 
The present error on $\alpha_s(M_Z^2)$ at NNLO of $\sim 0.0012$ is at the margin
of the present theory and systematics errors. Different known theoretical uncertainties, cf. also 
Eqs.~(\ref{eq:al1}--\ref{eq:al2}), are of the order of $0.0007$, the quoted $1 \sigma$ error of the world 
average. The systematics of the different extractions of $\alpha_s(M_Z^2)$ has 
to be understood in even more detail in the future. The current values of $\alpha_s(M_Z^2)$ 
obtained from precision deep-inelastic scattering data disfavor the unification of forces, 
even in the supersymmetric extension of the Standard Model. However, the picture may change 
soon, with new findings at the LHC. 
\begin{center}
\begin{figure}[t]
\hspace*{1cm}
\psfig{file=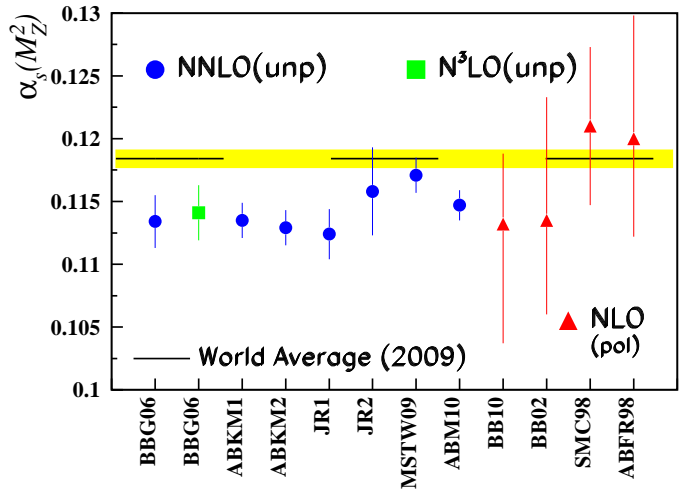,width=3.5in}
\caption[]{
The strong coupling constant $\alpha_s(M_Z^2)$ from different DIS
measurements, at NNLO and N$^3$LO, Eqs.~(\ref{eq:al1}--\ref{eq:al3}), 
and at NLO in the polarized case, {cf.~\cite{Blumlein:2010rn,Bluemlein:2002be}}. 
The yellow band describes
the weighted average of a wide range of $\alpha_s(M_Z)$ {measurements 
\cite{Bethke:2009jm}};~{Ref.~\cite{Blumlein:2010rn}}.
}
\label{aba:fig2}
\end{figure}
\end{center}
\vspace*{-7mm}

\vspace*{-3mm} 
\section{Main Inclusive Cross Sections at Hadron Colliders}\label{sec5}

\vspace*{1mm}\noindent
The accuracy for the parton distribution functions reached can now be applied
to derive precision predictions for inclusive hadronic observables, such as the
Drell-Yan cross section, the $W^{\pm}, Z$-boson, the $t\overline{t}$- and Higgs-boson
production cross sections at NNLO. Detailed analyses have been given 
{in~\cite{Alekhin:2009vn,JimenezDelgado:2008hf,JimenezDelgado:2009tv,Martin:2009iq,Alekhin:2010iu}}. For all 
these quantities at 
least
these corrections are necessary. The Drell-Yan cross section and  the $W^{\pm}, Z$-boson
production cross sections are, furthermore, used as `standard candle' processes to measure
the collider luminosity. They have therefore to be known as precisely as possible. As an example 
we show in Figure~4 predictions on the inclusive Higgs boson production at hadron colliders.
Within the present accuracy of the parton distribution functions the 
predictions~\cite{Alekhin:2009vn,Martin:2009iq} still show some differences, which are likely related
to the different gluon distributions at low scale and the values of $\alpha_s(M_Z^2)$. The predictions
agree for $s \sim M_Z$ but differ for higher mass scales.

\begin{figure}
\hspace*{1.7cm}
\psfig{file=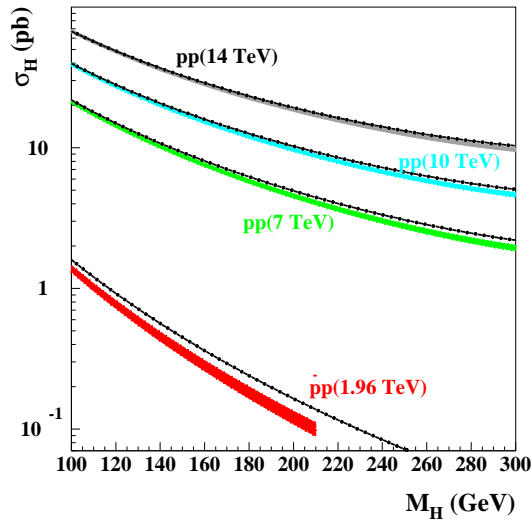,width=3in}
\caption[]{Inclusive Higgs boson production cross section at TEVATRON and the LHC. The bands denote
the parton distribution uncertainty of {ABKM09~\cite{Alekhin:2009vn}}; the lines correspond to 
{MSTW08~\cite{Martin:2009iq}}; from {Ref.~\cite{Alekhin:2009vn}}. 
}
\label{aba:fig3}
\end{figure}

\section{Higher Loop Integrals and Mathematics}\label{sec6}

\vspace*{1mm}\noindent
The computation of higher order loop integrals is still a difficult task, even 
in the massless case or in the presence of a single mass scale. At three-loop 
(and higher) orders this both applies to the zero-- and single--scale problems 
calculated at present. At the one hand, the results expected have a rather 
simple structure. On the other hand, a growing multitude of diagrams which contain  
more and more difficult structures have to be computed. Obviously, an enormous 
part of intermediary results simply cancels. It is, however, difficult to let
cancel these contributions at a rather early stage of the computation, or to even
widely avoid that they occur from the very beginning. This is one of the central 
problems of all present calculations. Gauss' 
theorem~\cite{LAGRANGE,*Chetyrkin:1980pr,*Laporta:2001dd} allows to express Feynman 
diagrams to a set of master integrals, which finally have to be computed for
zero scale quantities. Also in case of single scale quantities, e.g. given by
a Mellin variable $N$, one may obtain similar recursions. This method, simplifying
the calculation, may lead to a large number of terms being of higher 
complexity than those finally appearing in the results. This property seems to
be in common with different other approaches, which in the first place appear
to simplify the calculation technically, like  Mellin-Barnes~\cite{MB1a,*MB1b,*MB2} 
integrals or multinomial expansion, since they lead to rather elementary 
Feynman parameter integrals. Another approach, cf. {e.g.~\cite{Bierenbaum:2007qe,Bierenbaum:2008yu,
Ablinger:2010ha,Brown:2008um}}, consists in
evaluating the individual Feynman parameter integrals without applying intermediary 
simplifying methods, i.e. mapping them directly to the analytic mathematical structure 
they represent. This method is more demanding but will finally lead us to a deeper
understanding of the objects we deal with. For single scale single integrals integrals
and one mass,  the 2-loop integrals and simpler 3-loop integrals can be represented in terms of
generalized hypergeometric functions and extensions thereof, like Appell 
{functions cf.~\cite{Bailey,*Slater}}. An important issue in integration is
a clear definition of the target space and the knowledge of the relations of its elements.
For the Feynman parameter integrals discussed above, shuffle-- and {Hopf--algebras~\cite{
Kreimer:1997dp,*Connes:1998qv,*Broadhurst:1998ij,*Weinzierl:2003ub}}
play a central role, along with Poincar\'{e} iterated integrals over specific
alphabets. Feynman integrals are almost always {periods~\cite{Bogner:2007mn,*Bogner:2010kv}}.
In this context, the simplest structures are multiple zeta values, 
{cf.~\cite{Blumlein:2009cf}}, nested harmonic {sums \cite{Blumlein:1998if,*Vermaseren:1998uu}}, 
harmonic {polylogarithms~\cite{Remiddi:1999ew}}, and generalized harmonic {sums \cite{GON,Moch:2001zr,ABS}}. 
Feynman parameter integrals will coin new classes of higher transcendental functions going to higher
and higher order, which have to be studied to perform future precision calculations
in an efficient way. Their evaluation is intimately connected to modern summation technologies, like {{\tt 
SIGMA}~
\cite{Xsigma}}, and efficient algorithms to establish and solve the associated recurrences of 
both large order and {degree \cite{Blumlein:2009tj,Ablinger:2010ha}}.
All this will require high performance computer algebra written using
highly efficient languages like {{\tt FORM}~\cite{Vermaseren:2000nd}}, nearly 50 years after {\tt 
SCHOONSCHIP}
was {introduced~\cite{Veltman:1963}}, and the investment of many
CPU years, however, at an even more involved level than considered today. Structures of Feynman 
parameter integrals, e.g. on the level of multiple zeta values, form an interesting recent field 
in mathematics, {cf.~\cite{CARTIER,ZUDILIN,Brown:2008um}}, which is also related to irrationality proofs 
of the basis elements spanning these quantities. Here we face
a new era of a tight symbiosis between theoretical physics and modern mathematics, which is regarded to be 
very essential.
\section{Conclusions}\label{sec7}

\vspace*{1mm}\noindent
Quantum Chromodynamics was a great discovery. With it Murray Gell--Mann completed
the revolution of the strong interactions started in the early 1960ies with the 
introduction of the quarks. During the last 37 years
computations grew to a precision of $O(1\%)$ for inclusive 
quantities, which are described at  3--loop, and partly at 4--loop, level,
moving the frontiers of Quantum Field Theory to breathtaking new horizons.
The running of the strong coupling constant is understood in great detail,
despite in different classes of analyses still values of 
$\alpha_s(M_Z^2)$ are found which differ by experimental and 
theoretic systematic effects, being partly yet unknown. To determine
$\alpha_s(M_Z^2)$ at the level of its present statistical accuracy of $\sim 
1\%$ further studies and even higher order calculations are required for some 
of the processes. The QCD improved quark-gluon parton model works impressively
well at short distances - a clear triumph of Quantum Chromodynamics and proof
that quarks and gluons, although being confined, are basic building blocks of matter.
Without them the Standard Model would suffer from anomalies~\cite{Adler:1969gk,*Bell:1969ts,
*Bouchiat:1972iq} and not form a Quantum Field Theory.

During the last two decades the methods of lattice QCD steadily improved,
in particular concerning the systematic errors involved. We therefore expect
precision computations both on $\Lambda_{\rm QCD}$ and a series of moments
of unpolarized and polarized parton densities in the near future. The results
of these calculations ab initio can then be compared to the precision extractions
discussed based on precision experimental data and higher order perturbative
calculations. A final question concerns the unification of the three forces
of the Standard Model, given what we know at low scales at present. Future 
discoveries, perhaps at the LHC, will lead to a clarification here. 

\noindent
{\bf Acknowledgment.}~I would like to thank the organizers of the conference
for invitation. In a conversation with M. Gell--Mann it turned out that we 
have a second common interest: the physics of the direction of time. Everywhere 
in physics it runs forward, for all we know.

\vspace*{-3mm}
\bibliographystyle{ws-procs975x65}

\end{document}